\documentclass[graybox]{svmult}

\usepackage{mathptmx}       
\usepackage{helvet}         
\usepackage{courier}        
\usepackage{type1cm}        
\usepackage{makeidx}         
\usepackage{graphicx}        
\usepackage{multicol}        
\usepackage[bottom]{footmisc}

\makeindex             

\begin{document}

\title*{The effects of supernovae on the dynamical evolution of binary stars and star clusters}
\author{Richard J. Parker}
\institute{Richard J. Parker \at Astrophysics Research Institute, \newline Liverpool John Moores University, \newline 146 Brownlow Hill, \newline Liverpool L3 5RF, \newline United Kingdom. \newline \email{R.J.Parker@ljmu.ac.uk}}
%
%
\maketitle

\abstract{In this chapter I review the effects of supernovae explosions on the dynamical evolution of (1) binary stars and (2) star clusters. \\ \newline (1) Supernovae in binaries can drastically alter the orbit of the system, sometimes disrupting it entirely, and are thought to be partially responsible for `runaway' massive stars -- stars in the Galaxy with large peculiar velocities. The ejection of the lower-mass secondary component of a binary occurs often in the event of the more massive primary star exploding as a supernova. The orbital properties of binaries that contain massive stars mean that the observed velocities of runaway stars (10s -- 100s km\,s$^{-1}$) are consistent with this scenario. \newline (2) Star formation is an inherently inefficient process, and much of the potential in young star clusters remains in the form of gas. Supernovae can in principle expel this gas, which would drastically alter the dynamics of the cluster by unbinding the stars from the potential. However, recent numerical simulations, and observational evidence that gas-free clusters are observed to be bound, suggest that the effects of supernova explosions on the dynamics of star clusters are likely to be minimal.}

\section{Introduction}
\label{intro}

Binary stars and star clusters are both ubiquitous outcomes of the star formation process. The collapse of a star-forming core usually transports angular momentum to the outskirts, and the fragmentation results in a binary star system. Furthermore, protostellar discs are prone to fragmentation, which produces a binary system, and binary stars can also form through capture \cite{Goodwin07,Moeckel10}. On larger scales, the collapse and fragmentation of Giant Molecular Clouds results in groupings of stars, a fraction of which then become bound star clusters \cite{Lada03,Kruijssen12b}. 

Massive stars are more likely to be found in binaries than their lower-mass counterparts, whereas most star clusters fully sample the stellar Initial Mass Function \cite{Bastian10} and therefore contain massive stars. In this chapter, I will discuss the potential effects of superovae explosions on the dynamical evolution of both binaries and star clusters.

\section{Binary stars}
\label{sec:1}

The idea that supernovae \index{supernova} can directly affect the dynamical evolution of binary star \index{binary star} systems -- two stars orbiting a common centre of mass -- was first advocated by Zwicky in 1957 \cite{Zwicky57} and discussed in detail in two papers by Blaauw \cite{Blaauw61} and Boersma \cite{Boersma61}. These authors were attempting to explain the motion of massive stars (O- or B-type \index{massive stars!O-type star} \index{massive stars!B-type star} stars with masses $>$8M$_\odot$) in the Galaxy travelling with very high peculiar velocities, \index{peculiar velocity} i.e. velocities that differ with respect to the local frame of reference. These fast-moving massive stars were dubbed `runaways' \index{runaway star} as they could in some instances be traced back to star-forming regions \index{star forming region} and it was hypothesised that some mechanism had ejected them from these regions. 

It is possible that close encounters with other stars can lead to the ejection of massive stars \cite{Leonard90}, but this scenario has two immediate caveats: 

1) Firstly, ejecting stars of masses $>$8M$_\odot$ requires an interaction with a star at least as massive, if not more so than the ejected star. This comes from the theory of three-body dynamics, where a star can be replaced in a system only if the incoming star has a signifcantly higher energy than the binding energy of the system. If more than three stars are involved in the interaction, this constraint is relaxed somewhat, as the dynamics becomes more chaotic and/or stochastic.     

2) Regardless of the number of stars involved, the initial stellar density required to facilitate an interaction that would eject a massive star is likely to be well in excess of 1000\,stars\,pc$^{-3}$ \cite{Allison11,Oh15}. Due to their inherently peculiar velocities, runaway OB stars can often be traced back to their natal star-forming regions. These star-forming regions are often so-called OB associations \index{OB association}, which have much lower stellar densities than those required to dynamically eject massive stars. It is possible that OB associations were much more dense in the recent past \cite{Kroupa01a}, although recent observations suggest that they are more likely to have formed with the same spatial and kinematic properties observed today \cite{Wright14}. \\

Given that not all runaway OB stars are likely to have originated in very dense stellar environments, some other disruption mechanism must be responsible for producing them. We will first consider the general effects of stellar evolution (i.e.\,\,mass-loss) on the dynamical evolution of a binary star system, before discussing the effects of rapid, or instantaneous mass-loss due to a supernova.

\subsection{Evolution of binaries due to gradual mass loss}

If one of the stars in a binary system loses mass, then how will this affect the orbital configuration of the binary? According to Kepler's third law, \index{Kepler's third law} the period and semimajor axis are related to the component masses of the system thus:
\begin{equation}
P^2 = \frac{4\pi^2a^3}{G(m_1 + m_2)},
\end{equation}
where $P$ is the period of the system\index{binary star!period}, $a$ the semimajor axis,\index{binary star!semimajor axis} $G$ the gravitational constant and $m_1$ and $m_2$ the respective masses of the stars. The binding energy of the system is
\begin{equation}
E_{\rm bind} = -\frac{Gm_1m_2}{2a}.
\end{equation} 
This binding energy \index{binding energy} is equivalent to the total energy of the system (potential energy plus kinetic energy). We can write this as the energy per unit reduced mass, $E_{\rm red}$, where the reduced mass is
\begin{equation}
\mu = \frac{m_1m_2}{(m_1 + m_2)},
\end{equation} 
and so the binding energy per unit reduced mass is
\begin{equation}
E_{\rm red} = -\frac{G(m_1 + m_2)}{2a}.
\label{ebindred}
\end{equation}
In order to assess the effects of mass loss on the orbit of the binary we write the above equation in terms of the semimajor axis:\index{binary star!semimajor axis}
\begin{equation}
a = -\frac{G(m_1 + m_2)}{2E_{\rm red}}.
\label{sma}
\end{equation}  
We can make a differential equation of Eqn.~\ref{sma} to study the changes in $a$ and $E_{\rm red}$:
\begin{equation}
\frac{\delta a}{a} = \frac{\delta(m_1 + m_2)}{m_1 + m_2} - \frac{\delta E_{\rm red}}{E_{\rm red}}.
\label{change}
\end{equation}
Let us consider the kinetic and potential energies of the binary system:
\begin{equation}
E_{\rm red} = T_{\rm red} + \Omega_{\rm red},
\label{en_term} 
\end{equation}
where $T_{\rm red}$ and $\Omega_{\rm red}$ are the kinetic and potential energy per unit reduced mass, respectively. The kinetic energy term is: \index{kinetic energy}
\begin{equation} 
T_{\rm red} = \frac{1}{2}V^2,
\end{equation} 
where $V$ is the velocity of the secondary star, $m_2$ at a given point in the orbit. The potential energy term is: \index{potential energy}
\begin{equation}
\Omega_{\rm red} = -\frac{G(m_1 + m_2)}{r}, 
\end{equation}
where $r$ is the magnitude of the radius vector. We can now write Eqn.~\ref{en_term} as
\begin{equation}
E_{\rm red} = \frac{1}{2}V^2 - \frac{G(m_1 + m_2)}{r}.
\end{equation}
In order to compute the value of $\delta E_{\rm red}$ we need to consider the change in mass with time for the two component stars, using the differential operator $\delta/dt$. $\delta V/dt = 0$ and $\delta r/dt = 0$, hence:
\begin{equation}
\delta E_{\rm red} = \delta\Omega = -\frac{G\delta m_1}{r}. 
\end{equation}
Although it appears that $\delta E_{\rm red}$ depends on the exact point in the orbit $r$, we can assume that changes in the orbit occur slowly, and that only the time average values are important:
\begin{equation}
\delta E_{\rm red} \equiv \left<\delta E_{\rm red}\right> = -G\delta m_1 \left<\frac{1}{r}\right>.
\end{equation}
Because we are averaging over all points in the orbit, we have:
\begin{equation}
\left<\frac{1}{r}\right> = \frac{1}{a},
\end{equation}
and so 
\begin{equation}
\delta E_{\rm red} \equiv \left<\delta E_{\rm red}\right> = -\frac{G\delta m_1}{a}.
\label{deltae}
\end{equation}
Dividing Equations~\ref{deltae}~by~\ref{ebindred}, we obtain
\begin{equation}
\frac{\delta E_{\rm red}}{E_{\rm red}} = \frac{2\delta m_1}{m_1 + m_2}.
\label{deltaeav}
\end{equation}
Finally, assuming the secondary star $m_2$ is not losing mass, then we place Eqn~\ref{deltaeav}~into~\ref{change} to obtain 
\begin{equation} 
\frac{\delta a}{a} = - \frac{\delta m_1}{m_1 + m_2}.
\label{change_fin}
\end{equation}
Inspection of Eqn.~\ref{change_fin} shows that as the primary component loses mass ($\delta m_1$ is negative), the semimajor axis \index{binary star!semimajor axis} of the binary orbit increases. A similar analysis can also be performed to show that the period of the binary will also increase \cite{Huang56}.

If the mass-loss from one or both components causes the binary's semimajor axis to increase, here are two implications for the future evolution of the binary system. Firstly, the overall binding energy is reduced, which makes in more susceptible to subsequent dynamical disruption from interactions with passing stars, especially if the binary is still in its (dense) natal star-forming region. 

Secondly, the increase in period and semimajor axis leads to a decrease in orbital velocity. As well as the overall lower binding energy, the binary is more prone to encounters where a passing star can exchange into the binary (and eject one of the original stars). 

\subsection{Evolution of binaries due to rapid/instant mass loss}

The above considerations apply to a gradual loss of mass in the binary, which describes the dynamical evolution of the system before any supernova. \index{supernova} During the supernova, the more massive component loses mass almost instantaneously, and so the effect on the dynamical evolution of the system is quite different to the slow mass-loss case described above.

Let us re-write Eqn.~\ref{en_term} and drop the subscript `red' (the terms still however, refer to the energy-per-reduced mass) and instead replace them with `o', which refers to the total energy, $E_{\rm o}$, kientic energy $T_{\rm o}$ \index{kinetic energy} and potential energy $\Omega_{\rm o}$ \index{potential energy} \emph{before} mass loss:
\begin{equation}
E_{\rm o} = T_{\rm o} + \Omega_{\rm o}. 
\end{equation}
The potential energy, $\Omega_{\rm o}$ is 
\begin{equation}
\Omega_{\rm o} = -\frac{G(m_1^{\rm o} + m_2)}{R_{\rm o}}, 
\end{equation}
where $m_1^{\rm o}$ is the mass of the primary star immediately before the supernova, and $R_{\rm o}$ is the radius of the orbit (assumed to be circular). The kinetic energy changes slowly as the orbit reacts to the sudden mass-loss, but the potential energy changes immediately. 

The mass of the primary (i.e. supernova) decreases instantaneously by a factor $\epsilon$, so the new mass of the primary is $\epsilon m_1^{\rm o}$. The corresponding new potential energy,  $\Omega_{\rm n}$ is thus:
\begin{equation}
\Omega_{\rm n} = -\frac{G(\epsilon m_1^{\rm o} + m_2)}{R_{\rm o}}, 
\end{equation} 
and the change in potential energy is 
\begin{eqnarray}
\Delta\Omega = \Omega_{\rm n} - \Omega_{\rm o} \nonumber\\
\Delta\Omega = \frac{Gm_1^{\rm o}}{R_{\rm o}}(1 - \epsilon).
\end{eqnarray} 
The new total energy, $E_{\rm n}$, is therefore:
\begin{equation}
E_{\rm n} = T_{\rm o} + \Omega_{\rm o} + \frac{Gm_1^{\rm o}}{R_{\rm o}}(1 - \epsilon).
\label{en_term_new}
\end{equation}
Assuming the orbit is circular ($a = R_{\rm o}$), the total energy of the original system before mass-loss ($E_{\rm o} = T_{\rm o} + \Omega_{\rm o}$) can be written:
\begin{equation}
E_{\rm o} = -\frac{G(m_1^{\rm o} + m_2)}{2R_{\rm o}}.
\end{equation}
If we substitute this into Eqn.~\ref{en_term_new}, we have 
\begin{equation}
E_{\rm n} = -\frac{G(m_1^{\rm o} + m_2)}{2R_{\rm o}} + \frac{Gm_1^{\rm o}}{R_{\rm o}}(1 - \epsilon),
\end{equation}
which can be re-written as:
\begin{equation}
E_{\rm n} = \frac{G}{2R_{\rm o}}(m_1^{\rm o} - 2\epsilon m_1^{\rm o} - m_2).
\end{equation}
This equation describes the new energy of the binary system following the reduction in mass of the component that explodes as a supernova. A closed orbit (i.e. a bound binary system) has a total energy $E < 0$, whereas an open or hyperbolic orbit has a total energy $E > 0$. Therefore, in order to release the secondary component from the system, 
\begin{equation}
m_1^{\rm o} - 2\epsilon m_1^{\rm o} - m_2 > 0,
\end{equation}
or the reduction factor $\epsilon$ (recall that this is the ratio of post-supernova to pre-supernova mass of the exploding star) must fulfill the following inequality:
\begin{equation}
\epsilon < \frac{m_1^{\rm o} - m_2}{2m_1^{\rm o}}.
\end{equation}  
It is immediately apparent that $\epsilon$ must be less than 1/2 and this is only in the limit that $m_1^{\rm o} >> m_2$; in reality, the masses of the two components may actually be comparable (see Section~\ref{demographics} on binary demographics)  and so the reduction factor $\epsilon$ would need to be even lower.

For an O-type star \index{massive stars!O-type star} of initial mass $m_1 = 25$M$_\odot$, the typical pre-supernova mass $m_1^{\rm o}$ can be anywhere in the range  $m_1^{\rm o}= $10 -- 17M$_\odot$ \cite{Meynet03}. Following the supernova, the remaining compact object \index{compact object} (a neutron star) will have a mass of $m_{1,{\rm co}} = 1.4$M$_\odot$, implying a reduction factor of $\epsilon \sim 1/10$. Stars with initial mass  $m_1 > 25$M$_\odot$ will typically form a black hole with a higher mass -- $m_{1,{\rm co}} = 5$M$_\odot$, but still have a similar pre-supernova mass  ($m_1^{\rm o}= $10 -- 17M$_\odot$)  and so the reduction factor could be closer to $\epsilon \sim 1/2$.

The above considerations -- although simplified through various assumptions -- suggest that most supernovae in binaries result in the release of the secondary component. 

Detailed numerical simulations have shown that the sudden mass-loss resulting from a supernova can often unbind the secondary star from the binary system \cite{Boersma61,Tauris98,Zwart00}. It is also possible, especially in the case of assymetric mass-loss from the primary component, that the velocity kick resulting from the supernova explosion can eject both the secondary star $m_2$, \emph{and} the resulting compact object $m_{1,{\rm co}}$, \index{compact object} from the host star-forming region. 

\subsection{Binary demographics}
\label{demographics}

 In addition to the arguments above a pertinent question to ask is what fraction of massive stars may be affected by this mechanism? To answer this, we need to understand the fraction of massive stars that are in binary systems, and the distributions of their orbital semimajor axes and mass ratios. \index{massive stars}

The binary fraction of stars, $f_{\rm bin}$ is \index{binary fraction}
\begin{equation}
f_{\rm bin} = \frac{B + T + \ldots}{S + B + T + \ldots},
\end{equation}
where $S$, $B$, and $T$ are the numbers of single, binary or triple systems, respectively, in either a star-forming region, or a volume--limited region of the Galactic disc. For binary stars where the primary component is of similar mass to the Sun, the binary fraction is roughly 0.5 \cite{Raghavan10}. The orbital separation distribution can be approximated by a log-normal with a peak at $\sim$50\,au and a variance ${\rm log}_{10} a = 1.68$, which means that the binaries can have semimajor axes anywhere between 10$^{-2}$au and 10$^5$au. The distribution of mass ratios \index{binary mass ratio} ($q = m_2/m_1$) is flat, i.e. a binary is equally likely to have a secondary component whose mass is 10 per cent of the mass of the primary, as opposed to a secondary component whose mass is equal to the primary mass \cite{Reggiani11a}. 

Unfortunately, the binary statistics for the most massive stars (i.e.\,\,those that may contain a supernova that will affect the dynamics of the binary system) are hampered by the simple fact that very few binaries that contain massive stars exist. Recent research \cite{Mason09,Sana13} has gone some way to addressing this problem, mainly by targeting binaries containing massive stars in young massive star clusters \index{star cluster}. These environments are so young that the massive stars have not yet evolved (or gone supernovae), and because they are so massive they contain many massive stars.

Nevertheless, for binary systems that do contain massive stars it is apparent that their properties (overall fraction and orbital parameters) are very different to those of binaries where a solar-mass star is the primary component. A general consensus is that the binary fraction in young massive clusters is between 0.7 and 1.0 \cite{Mason09,Sana13}, and that their semimajor axis \index{binary star!semimajor axis} distribution is markedly different from lower-mass binaries, in that there is an excess of very close ``spectroscopic'' systems. Their semimajor axes follow an Opik \cite{Opik24} distribution, i.e. flat in log-space between 0 and 50\,au. Furthermore, the mass ratio distribution appears to favour equal-mass systems ($q \sim 1$), meaning that an O-type star is more likely to be paired with either another O-type star, or a very massive B-type star. \index{massive stars!O-type star} \index{massive stars!B-type star}.

\subsection{Properties of runaway stars}

How do the binary properties of massive stars affect the expected velocity distribution of the runaway stars? \index{runaway star} According to calculations by \cite{Boersma61,Tauris98,Zwart00}, the velocity of the runaway star (the secondary component in the binary) is proportional to the orbital velocity of the binary, $V_{\rm orb}$:
\begin{equation}
V_{\rm 2,run} = \sqrt{1 - 2\frac{m_1^{\rm o}/m_{1,\rm co} + m_2/m_{1,\rm co}}{(m_1^{\rm o}/m_{1,\rm co})^2}}V_{\rm orb},
\end{equation}
which for a supernova where the compact object that forms is a neutron star ($m_{1,\rm co} \sim 1$M$_\odot$), 
\begin{equation}
V_{\rm 2,run} \simeq \sqrt{1 - 2\frac{m_1^{\rm o} + m_2}{(m_1^{\rm o})^2}}V_{\rm orb}.
\end{equation}
As before, $m_1^{\rm o}$ is the mass of the primary star immediately before the supernova, and $m_2$ the mass of the secondary (runaway) star.

The orbital velocity, $V_{\rm orb}$, at the point of the supernova \index{supernova} is
\begin{equation} 
V_{\rm orb} = \sqrt{\frac{Gm_1^{\rm o}m_2}{(m_1^{\rm o} + m_2)a}},
\end{equation}
where $m_1^{\rm o}$ and $m_2$ are the pre-supernova masses of the primary and secondary stars, respectively, and $a$ is the semimajor axis. 

Assuming the observed Opik distribution, then the maximum semimajor axis is 50\,au. If we assume both stars have masses of 20M$_\odot$, then the velocity of the runaway after the supernova is 13km\,s$^{-1}$. If the stars are both 40M$_\odot$, then the velocity is slightly higher, at 19km\,s$^{-1}$. Adopting a much smaller semimajor axis (e.g. 1au) results in predicted velocities of order 100km\,s$^{-1}$. However, massive stars that explode as supernovae lose a substantial fraction of their mass before the explosion. Therefore, the pre-supernovae masses $m_1^{\rm o}$ and $m_2$ are likely to be in the region of 10 -- 17M$_\odot$. Even with these values, a binary with initial semimajor axis 1\,au is likely to impart a velocity of 100km\,s$^{-1}$ onto the secondary star after the supernova. 

\subsection{Dependence on star formation environment?}

In determining the fraction of binary stars that could be disrupted by supernovae we must implicitly assume that each star forming region \index{star forming region} has the same binary star \index{binary star} properties -- a so-called `Universal population' \cite{Kroupa95a}. However, it is far from established that the binary fraction, \index{binary fraction} and orbital characteristics are independent of environment. For example, in star-forming regions near the Sun the populations of binaries have been argued to be identical at birth, and observed differences are merely due to different amounts of dynamical processing. In this picture, the overall binary fraction and the semimajor axis distributions in two different regions can be altered to different degrees by the regions being born with different stellar densities (a high density region destroys more binaries compared to a lower-density region). 

This is important for determining the fraction of binary systems that could be disrupted via supernovae. Dynamical interactions in dense star clusters \index{star cluster} act over crossing times -- the distance travelled by stars in the cluster divided by the velocity; often taken to be the radius and average velocity. In very dense regions the crossing time can be as low as 0.1Myr, i.e. far shorter than the stellar evolution timescales for the most massive stars. Therefore, binaries containing massive stars could be disrupted through dynamical interactions with other massive stars before any supernova explosion.

Alternatively, other authors argue that based on constraints from the spatial distributions of stars in young star clusters, the initial binary population may vary between regions (possibly as a result of a different environment during the star formation process, \cite{Horton01,King12a}). Young star-forming regions exhibit a large degree of spatial substructure\cite{Cartwright04}, and this is erased on order of a few crossing times due to dynamical interactions. Therefore, the amount of substructure in a star-forming region places an upper limit on the amount of dynamical interactions that may have taken place\cite{Parker14e}. If the binary population does vary as a function of environment, then different star-forming regions may produce runaways due to supernovae in different proportions. 

\section{Star clusters}
\label{sec:2}

\subsection{What is a star cluster?}

Before discussing the effects of supernovae on the dynamical evolution of star clusters, \index{star cluster} it is worth a digression on the exact definition(s) that exist for classifiying clusters. In undergraduate courses we all learn that Globular Clusters \index{star cluster!Globular cluster} are old ($>$Gyr), massive ($>10^5$ M$_\odot$) dense stellar systems that are believed to have formed early in the evolution of the Universe. Their younger, less massive counterparts are often referred to as Open Clusters, \index{star cluster!Open cluster} and these have ages between 10 Myr -- 1 Gyr, masses between $100 - 10^4$ M$_\odot$ and fairly low densities (typically$ <10^3$ M$_\odot$pc$^{-3}$).

However, Type II supernovae occur in stars with progenitor masses $>$ 8M$_\odot$, and very massive O-type stars \index{massive stars!O-type star}($>$20M$_\odot$) usually evolve (and explode as supernovae) on timescales less than 10Myr. At ages less than 10Myr, supernovae are therefore not found in Open or Globular clusters but rather in regions of recent and/or ongoing star formation. 

As discussed in the previous section, a significant fraction of massive stars are observed in extended complexes (often several to 10s pc across) of star formation referred to as OB associations\cite{Blaauw64}. \index{OB association} How these OB associations form is currently the subject of debate. One school of thought is that they form essentially in situ, that is to say the the current observed density and spatial distribution of the stars (and velocity distribution, if measured) is very similar to that set by the initial conditions of star formation in the region in question. The second hypothesis is that OB associations are the expanded remnants of initially very compact (radii less than 1pc), dense groups of stars.

Young stars are typically found still embedded in the giant molecular clouds (GMCs) \index{Giant Molecular Cloud, GMC} from which they form. Advances in infra-red detectors in the 1980s and 1990s meant that GMCs were studied in detail for the first time, and significant new groups of stars were discovered (see the review by Lada \& Lada,\cite{Lada03}). These were dubbed `embedded clusters' \index{embedded cluster} and until recently the prevailing view in the literature was that almost all stars (70 -- 90\%) formed in these compact, embedded clusters. 

However, recent observations suggest the picture may be somewhat more complicated. Alternative theories postulate the stars form in scale-free hierarchically self-similar spatial distributions \cite{Efremov95,Elmegreen01,Kruijssen12b}. Some star-forming regions may undergo a dense phase, where dynamical interactions dominate and a dense cluster forms, but it is unclear what fraction of stars experience a clustered phase in their lifetime.

Despite this, we can still think of the estimate of the fraction of stars that form in clusters from Lada \& Lada as an upper limit, as it is unclear which type of star-forming region (clustered versus non-clustered) contributes the most stars to the Galactic disc. 

A further observation of embedded and young clusters is a sharp drop-off in the numbers of clusters as a function of their age. After 10\,Myr, the number of clusters has decreased by 90\% when compared to the number of clusters with ages of 1\,Myr \cite{Lada03,Pellerin07}. This has been coined rather distastefully as ``infant mortality'' in the literature. The implication is that some mechanism(s) is destroying or disrupting the clusters as they age. 

\subsection{Star formation efficiency and gas expulsion}

Even before the discovery of a decrease in the number of embedded clusters with age, it had been postulated that some processes act to destroy star clusters. Star formation is typically an inherently inefficient process. Stars form from the collapse and fragmentation of giant molecular clouds, but usually only around 30\% of the gas mass is converted into stars. This means that 70\% of the mass remains in the form of molecular gas and the `star formation efficiency' is only 30\% \index{star formation efficiency}. 

Given that such a substantial fraction of the system mass is in the form of gas, any process that would act on the gas but not the stars may cause a significant change to the system in question. 

Stars appear to form with a mass distribution that is independent of environmental factors, such as the mass, density of gas and velocity motions in the gas. A more contentious issue is whether this mass distribution -- the `Initial Mass Function' \index{Initial Mass Function}, or IMF -- is also independent of the age of the universe at the point of formation. Nevertheless, in nearby star clusters the IMF is statistically very similar \cite{Bastian10} and follows a roughly log-normal distribution for stellar masses ranging from the hydrogen burning limit, 0.08M$_\odot$, to around 1M$_\odot$. At higher masses, the IMF follows a power-law slope of the form
\begin{equation} 
\frac{dN}{dM} \propto m^{-\alpha},
\end{equation} 
where $\alpha = 2.35$ - the so-called Salpeter slope \cite{Salpeter55}. 

If we visualise this distribution, we immediately see than the probability of selecting a low-mass star is much higher than the probability of selecting a high-mass star. If we think of stellar mass distributions as an abstract maths problem, we now need to populate a star cluster of a given stellar mass with stars drawn from the IMF until we reach the cluster mass \cite{Elmegreen06}. 

Therefore, on average a star cluster that contains several massive stars ($>$20\,M$_\odot$) will contain hundreds of other, low-mass stars. Massive stars evolve on timescales less than 10\,Myr, so we expect they will lose a significant fraction of their mass during the early stages of a star cluster. However, if the IMF has been well-sampled (i.e. the cluster is also populated with many other (low-mass) stars), then the fraction of mass lost by the cluster due to the massive stars evolving and exploding as supernovae is negligible in the first 10\,Myr. 

However, recall that the vast majority (70\% by mass) of gas is not converted into stars. Observations of H{\small II} regions \index{H{\small II} regions} -- the immediate vicinities of the most massive stars -- are often free of gas, which is interpreted as being due to the removal of gas by the intense stellar winds of the most massive stars \cite{Matzner02,Krumholz06}. The same effect is predicted to occur from the final supernova explosion. 

What is the effect of this gas removal on the dynamical evolution of the system? Let us consider a simplistic view of the binding energy, $E_{\rm bind}$ \index{binding energy} of a star of mass $m_\star$ in the system:
\begin{equation}
E_{\rm bind} = -\frac{GM_{\rm gas}m_{\star}}{2r},
\end{equation}
where $r$ is the distance between the star and the centre of mass of the gas, which has mass $M_{\rm gas}$. If a significant fraction of the gas mass is removed, then the binding energy of the individual stars will decrease.

If the system is initially in virial equilibrium, \index{virial equilibrium} then we have
\begin{equation}
2T + \Omega = 0,
\end{equation}
(where $T$ is the kinetic energy \index{kinetic energy} and $\Omega$ is the potential energy) \index{potential energy} and so
\begin{equation}
T = -\frac{\Omega}{2}.
\end{equation}
If the cluster experiences sudden mass loss due to instantaneous gas expulsion caused by the supernova, then the total cluster mass (stars + gas) is reduced by a factor of $\epsilon$. The potential energy becomes $\epsilon\Omega$ and we can write the total energy as:
\begin{equation}
E = \epsilon\Omega - \frac{\Omega}{2},
\end{equation}
or
\begin{equation}
E = (\epsilon - 0.5)\Omega.
\end{equation}
In order for the cluster to be unbound, the mean energy per star must be positive, which will happen if $\epsilon < 0.5$ (because the potential energy, $\Omega$ is negative). (This is analagous to the condition we derived for binary destruction by supernovae in the previous section.) Hence, for a virialised system where the stars and gas are well-mixed, supernovae can unbind the system if a significant fraction of the mass of the cluster (in this case the gas left over from star formation) is expelled. 

One can also show that in the case of rapid mass loss, the cluster will expand by a factor
\begin{equation}
\frac{R_f}{R_o} = \frac{1 - \epsilon}{1 - 2\epsilon},
\end{equation}
where $R_o$ and $R_f$ are the radii before and after instantaneous mass-loss \cite{Hills80}.

\subsection{Predictions from simulations}

If the binding energy of stars suddenly decreases due to a supernova, their velocities increase to the point that they are no longer bound to the centre of mass of the cluster. This has been shown in numerous theoretical and numerical works where the gas is modelled as a background potential \cite{Tutukov78,Lada84,Goodwin06,Baumgardt07}. 

In $N$-body simulations of star cluster evolution, it is possible to model the gas as a single potential, which is either static or is time-dependent (i.e. the potential can be made to decrease with time to mimic the depletion of gas). This technique implicitly assumes that the stars and gas are well-mixed but they do not interact apart from in a gravitational sense. 

As $N$-body simulations are computationally inexpensive, many calculations that include a background gas potential have been run by various authors. In general, they find that the effect of removing the gas is to unbind a significant fraction of the stellar mass. The action of unbinding this mass is to increase the velocity dispersion \index{velocity dispersion} of the stars. A system where most of the stellar mass is not bound is said to be `supervirial', \index{supervirial} and the system will have a higher velocity dispersion than the equivalent system that is in virial equilibrium. 

The physical outcome of a supervirial system  is for it to expand. When a star cluster expands the stars start to feel less of the cluster's influence and more of the tidal field of the Galaxy. This so-called `Jacobi radius' \index{Jacobi radius} is given by
\begin{equation}
r_J = D_G\left(\frac{M_{\rm cl}}{3M_G}\right)^{\frac{1}{3}},
\end{equation}
where $D_g$ is the Galactocentric distance, $M_G$ the mass of the Galaxy and $M_{\rm cl}$ is the mass of the cluster.

For a cluster of $\sim$1000\,M$_\odot$ at the position of the Sun in the Milky Way's potential, the Jacobi radius at which the tidal field of the galaxy dominates over the cluster is at around 7\,pc. Any star exceeding this radius is likely to escape the cluster and become a member of the Galactic disc.

Because this Jacobi radius is a function of cluster mass, the more members that are lost, the smaller the radius becomes. For this reason, clusters that are supervirial (expanding) are likely to dissolve more readily than those in virial equilibrium. The chances of a cluster surviving the gas removal phase therefore rest on the star formation efficiency \index{star formation efficiency} (i.e. how much mass remains as gas) and the efficiency of supernovae to expel the gas and unbind the cluster. 

\subsection{Observational evidence?}

Until recently, observations of star clusters appeared to corroborate this picture. The observed velocity dispersions \index{velocity dispersion} of many clusters were consistent with the clusters being in a supervirial state when compared with an estimate for the dispersion for the cluster to be in virial equilibrium \cite{Bastian06,Olczak08}. The virial velocity dispersion is estimated using 
\begin{equation}
\sigma_{\rm vir} = \sqrt{\frac{2GM}{\eta R}},
\label{virial_mass}
\end{equation}
where $G$ is the graviational constant and $M$ is the enclosed mass within a radius $R$. $\eta$ is the structure parameter, and typically is $\sim$10 for a dynamically evolved, centrally concentrated cluster. If $\sigma_{\rm vir}$ is significantly higher than the observed velocity dispersion, the region is said to be supervirial, or unbound. \index{supervirial}

However, estimates of the velocity dispersion in clusters had failed to account for the the orbital motion of binary stars. The fraction of stars that are in binary \index{binary fraction} systems is around 50\% in the solar neighbourhood, and is thought to be similar, if not higher in young star clusters. A substantial fraction of binaries have semimajor axes \index{binary star!semimajor axis} less than 1\,au, so-called spectroscopic binary stars. The component stars of these binaries have extremely fast orbital motions, which dominate the velocity of the entire binary system with respect to its motion around the centre of mass of the cluster itself. 

If the underlying velocity dispersion for the motion of stars can be approximated as a Gaussian, then the contribution of the orbital motion of binary stars both adds extra signal to the wings of the Gaussian, and widens it. The net effect is to make a measured velocity dispersion appear higher than it really is, and this gives the impression that the star cluster is undergoing supervirial expansion, when it is in fact in virial equilibrium \cite{Gieles10}.

\subsection{The effects on a realistic potential}

When the orbital motion of binary stars \index{binary star} is accounted for, most star clusters are observed in virial equilibrium \cite{Cottaar12b} \index{virial equilibrium}. Several clusters also appear to be devoid of gas, suggesting that the star formation efficiency \index{star formation efficiency} is either unusually high (and so the removal of gas has not affected the dynamics of the stars) or that the gas removal has not affected the dynamics.

Clues as to why removing a significant amount of the potential does not influence the subsequent dynamical evolution come from hydrodynamic simulations of the earliest stages of star formation. Typically, these simulations replace overdensities of gas with `sink particles' in order to speed up the calculation. Depending on the resolution limit of the simulation, sink particles are used to model either entire star clusters, or single stars. They are assigned a radius, where any gas crossing this is removed from the simulation and added to the mass of the sink. 

In simulations of star formation, several researchers \cite{Kruijssen12a,Dale12b} have shown that the locations where the sink particles form are almost entirely devoid of gas. The physical interpretation of this is that the local star formation efficiency is high, so most of the gas is converted to sinks, which are spatially arranged in groups. Therefore, the effects  on the dynamics of the stars of instantaneously removing this gas due to supernovae are virtually negligible, because the potential of the remaining gas is so small. If this argument scales up to clusters of 100 -- 1000\,M$_\odot$, then it is possible that the supernovae explosions of massive stars have no effect on the dynamical evolution of the cluster. Furthermore, any cluster that is gas free at ages of less than $\sim$5Myr will never be unbound due to supernovae, because supernovae only start to explode at these ages.

In summary, it appears that the effects of supernovae on the dynamics of star clusters are probably minimal, although this debate is currently ongoing in the literature.

\section{Conclusions}

I have discussed the effects of rapid mass-loss experienced by a bound binary star system due to a core-collapse supernova and shown that this often leads to the disruption of the system. Furthermore, the secondary, less massive star is often ejected at high velocity -- a so-called `runaway' star. Because binary systems which contain stars massive enough to explode as supernovae have a higher likelihood of containing another massive star of comparable mass, the runaway stars produced by supernovae are readily observable. In some cases they can be traced back to their natal star-forming regions, which places constraints on the birth environment of massive star binaries and the initial density of star formation.

The effects of supernovae on the dynamical evolution of star clusters are more modest as the total stellar mass of the star cluster does not significantly decrease during the supernova(e). However, because star formation is so inefficient, a significant fraction of the mass in a star cluster is thought to be in the form of gas left over from the star formation process. Many authors have argued that supernovae could remove this gas and therefore unbind the stars in the cluster, leading to the dissolution of the cluster. Recent observational and theoretical work has shown that the effects of this gas removal on the dynamical evolution of a star cluster are likely to be minimal.

\section{Acknowledgements}

I am grateful to John Bray for pointing out several typos in equations 24 and 25, which appeared in the original version of this manuscript. I acknowledge support from the Royal Astronomical Society in the form of a research fellowship.

\bibliographystyle{spphys}
 \bibliography{general_ref}

\printindex

\end{document}